\documentclass[twocolumn,superscriptaddress,preprintnumbers,amsmath,amssymb]{revtex4}
\usepackage{graphicx}
\usepackage{dcolumn}
\usepackage{bm}
\usepackage{hyperref}
\usepackage{amssymb,amsmath,amsthm,amsfonts}
\usepackage{bbm}
\usepackage{mathtools}
\usepackage{enumitem}
\usepackage{color}
\usepackage{textcomp}
\usepackage{gensymb} 
\usepackage{float}
\usepackage{soul}
\usepackage{booktabs}
\usepackage{graphicx}
\usepackage{siunitx}
\usepackage{multirow}
\usepackage{makecell}
\usepackage{draftwatermark}

\newcommand{\eq}[1]{(\ref{eq:#1})}
\renewcommand{\sec}[1]{\hyperref[sec:#1]{Section~\ref*{sec:#1}}}
\newcommand{\fig}[1]{\hyperref[fig:#1]{Figure~\ref*{fig:#1}}}
\newcommand{\tab}[1]{\hyperref[tab:#1]{Table~\ref*{tab:#1}}}

\global\long\def\l({\left(}
\global\long\def\r){\right)}

\newcommand{\comment}[1]{\textbf{\color{red}}}

\makeatletter
\newcommand{\vast}{\bBigg@{4}}
\newcommand{\Vast}{\bBigg@{5}}
\newcommand{\VVast}{\bBigg@{11.8}}
\makeatother
\begin{document}

\title{Efficient Arbitrary Simultaneously Entangling Gates on a trapped-ion quantum computer}

\author{Nikodem Grzesiak}
\email{grzesiak@ionq.co}
\affiliation{IonQ, College Park, MD 20740, USA}
\author{Reinhold Bl\"umel}
\email{blumel@ionq.co}
\affiliation{IonQ, College Park, MD 20740, USA}
\affiliation{Wesleyan University, Middletown, CT 06459, USA}
\author{Kristin Beck}
\affiliation{IonQ, College Park, MD 20740, USA}
\author{Kenneth Wright}
\affiliation{IonQ, College Park, MD 20740, USA}
\author{Vandiver Chaplin}
\affiliation{IonQ, College Park, MD 20740, USA}
\author{Jason M. Amini}
\affiliation{IonQ, College Park, MD 20740, USA}
\author{Neal C. Pisenti}
\affiliation{IonQ, College Park, MD 20740, USA}
\author{Shantanu Debnath}
\affiliation{IonQ, College Park, MD 20740, USA}
\author{Jwo-Sy Chen}
\affiliation{IonQ, College Park, MD 20740, USA}
\author{Yunseong Nam}
\email{nam@ionq.co}
\affiliation{IonQ, College Park, MD 20740, USA}





\begin{abstract}





Efficiently entangling pairs of qubits is essential to fully harness the power of quantum computing. Here, we devise an exact protocol that simultaneously entangles arbitrary pairs of qubits on a trapped-ion quantum computer. The protocol requires classical computational resources polynomial in the system size, and very little overhead in the quantum control compared to a single-pair case. We demonstrate an exponential improvement in both classical and quantum resources over the current state of the art. We implement the protocol on a software-defined trapped-ion quantum computer, where we reconfigure the quantum computer architecture on demand. Together with the all-to-all connectivity available in trapped-ion quantum computers, our results establish that trapped ions are a prime candidate for a scalable quantum computing platform with minimal quantum latency.






\end{abstract}

\maketitle
Quantum computers are expected to solve certain computational problems of interest more efficiently than classical computers using state-of-the-art classical algorithms. Notable examples include integer factorization \cite{ar:Shor}, unsorted database search \cite{ar:Grover}, and quantum dynamics simulations \cite{ar:Feynman}. Multiple quantum computing platforms are under active development today. One of these platforms is the trapped-ion quantum information processor (TIQIP), which has demonstrated  ${}^{171}{\rm Yb}^+$ qubit coherence times in excess of 10 minutes \cite{ar:Yewang}, single-qubit gate fidelity of $99.9999\%$ \cite{Harty2014}, and two-qubit gate fidelity of $99.9\%$ \cite{Gaebler2016,Ballance2016}. Additionally, a TIQIP may leverage the all-to-all connectivity between ion qubits. The ability to directly apply a two-qubit gate to any pair of qubits provides TIQIPs an important advantage over other QIPs with limited connectivity \cite{ar:LinkePNAS}. 

While the current progress in TIQIP technology is remarkable, better quality quantum gates are needed to run longer quantum programs and still obtain reliable quantum computational results \cite{ar:IEEE}. The shortest quantum program known to date, expected to deliver scientifically meaningful discoveries, requires hundreds of thousands of quantum gates \cite{ar:HINT}. Therefore, to address quantum computational problems of broad interest, the two-qubit gate design in TIQIPs must be improved. An efficient procedure that simultaneously implements as many two-qubit gates as possible with the least amount of resources will thus accelerate the process of harnessing the power of universal, programmable quantum computers.

In this paper, we devise a new protocol that efficiently and simultaneously implements multiple two-qubit gates on a TIQIP. Using our efficient, arbitrary, simultaneously entangling (EASE) gates, arbitrary ion-qubit pairs, overlapping or not, can be entangled with programmable degrees of quantum entanglement. We implement EASE gates by modulating the amplitude of laser pulses that address individual ion qubits that comprise our scalable, general-purpose, programmable TIQIP, hosted at IonQ \cite{system2}. These new gates pave the way for efficient implementations of large-scale quantum algorithms on a TIQIP.

\section{Two-qubit gate on a trapped-ion quantum information processor}

\begin{figure*}[ht!]
\centering
\includegraphics[scale=0.39]{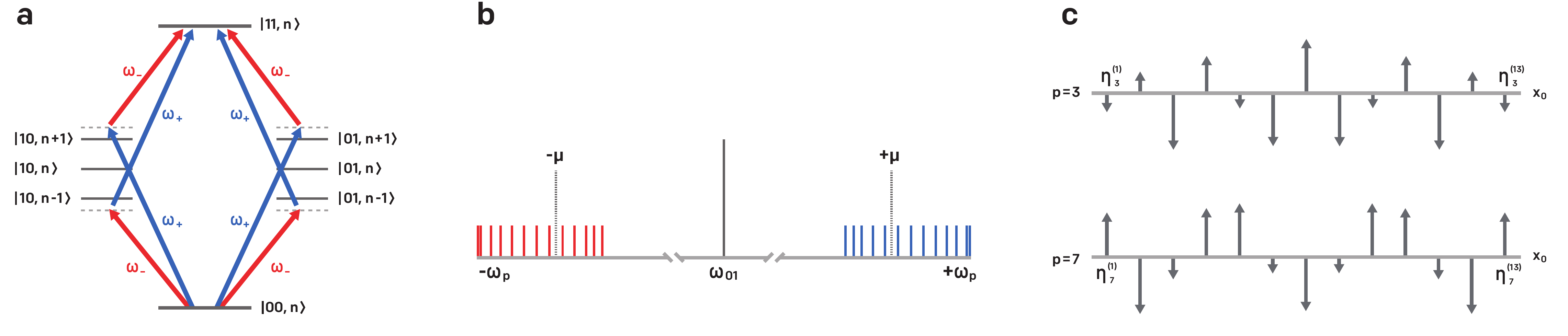} 
\caption{\textbf{Quantum dynamics of an EASE gate.} {\bf a} An energy-level diagram associated with the $p$th motional mode showing off-resonant red and blue sideband transitions that cause the desired two-qubit coupling between $|00\rangle$ and $|11\rangle$ quantum states. Here, $\omega_\pm = \omega_{01} \pm \mu$, and $n$ denotes the of motional state. {\bf b} Frequency spectrum of the motional modes $\omega_p$ of the ion chain centered around the carrier frequency $\omega_{01}$ that induces a single-qubit state transition. Symmetric detuning by frequency $\mu$ for red and blue sidebands is applied to the pulses that illuminate ions to induce the desired EASE gate. {\bf c} Motional-mode diagrams that show the geometric structure of the modes. The ion displacements from their respective equilibrium positions are proportional to the coupling strength $\eta_p^{(m)}$ between the different ions $m$ and the different modes $p$.}
\label{fig:apparatus}
\end{figure*}

The native two-qubit gate on our TIQIP is implemented according to the M{\o}lmer S{\o}rensen protocol \cite{ar:MS1,ar:MS2,ar:MS3}, which induces an effective XX-Ising interaction between a pair of qubits. The coupling between the computational states of the qubit pair is mediated by the motional modes of the linear ion chain stored in an ion trap. The evolution operator $\hat{U}$ that describes this operation is \cite{AM1}
\small
\begin{equation}
\label{eq:Evol}
\hat{U} = \exp\left[ \sum_{m=1}^{N} (\hat{\beta}^{(m)}-\hat{\beta}^{(m) \dagger} )\hat{\sigma}_x^{(m)} - i \sum_{n\neq m} \chi^{(m,n)} \hat{\sigma}_x^{(m)} \hat{\sigma}_x^{(n)}/4 \right]
\end{equation}
\normalsize
where $\hat{\beta}^{(m)} = i\sum_{p=1}^N \alpha_{p}^{(m)}(\tau)\hat{a}_p^\dagger$ (with motional mode index $p$, coupling strength $\alpha_{p}^{(m)}$ between ion $m$ and mode $p$, the $p$th motional-mode creation operator $\hat{a}_p^\dagger$ -- see \fig{apparatus} -- and the gate duration $\tau$) denotes the coupling between the computational state of qubit $m$ and the motional modes, $\hat{\sigma}_x^{(m)}$ is the Pauli-$x$ operator on the $m$th qubit, and $\chi^{(m,n)}$ denotes the degree of entanglement between qubits $m$ and $n$. To obtain a successful single-pair XX gate, we require that the first term in \eq{Evol} and all $\chi^{(m,n)}$ vanish, except for $\chi^{(m,n)}$ of the targeted ion pair $m,n$. Similarly, to implement EASE gates between freely chosen pairs of qubits with an arbitrary degree of entanglement for every pair, we require that 
\begin{enumerate}[label=(\Alph*)]
\item the first operator $\hat{\beta}^{(m)}$, which represents the coupling between motional modes of the ion chain and the computational states of the qubits, vanishes at the end of the evolution, and that
\item the second operator's coefficient $\chi^{(m,n)}$ either vanishes (if the ion pair $m,n$ is not to be entangled) or computes to a pre-specified degree of entanglement (if the pair is to be entangled).
 \end{enumerate} To satisfy conditions (A) and (B), we individually address participating ions with amplitude-modulated (AM) laser pulses \cite{system2}, where the modulation is performed by dividing the gate time $\tau$ into $N_\text{seg}$ equi-spaced segments and allowing the amplitude to vary from one segment to the next. 

Denoting the amplitude of the pulse $\Omega^{(m)}(t)$ applied to ion $m$ during segment $k$ as $\Omega_k^{(m)}$, the laser detuning from the carrier frequency as $\mu$ and the motional-mode frequencies as $\omega_p$, condition (A) implies, for all $m$ and $p$,
\begin{align}
\label{eq:condA}
&\alpha_{p}^{(m)}(\tau) = -\eta_p^{(m)} \int_0^\tau dt\, \Omega^{(m)}(t) \cos(\mu t) e^{i\omega_p t} = 0 \\ \nonumber
\mapsto
&\sum_{k=1}^{N_\text{seg}} \Omega_k^{(m)} \int_{(k-1)\tau/N_\text{seg}}^{k\tau/N_\text{seg}} dt\, \cos(\mu t) e^{i\omega_p t} = \hat{M} \vec{\Omega}^{(m)} = 0,
\end{align}
where $\eta_{p}^{(m)}$ denotes the coupling constant (Lamb-Dicke parameter) for qubit $m$ and mode $p$ (see also \fig{apparatus}), $\hat{M}$ is the matrix with elements that are the segmented integrals shown above, and $\vec{\Omega}^{(m)}$ is the vector of $\Omega_k^{(m)}$.
Likewise, in the segmented form, condition (B) implies
\small
\begin{align}
\label{eq:condB1}
\chi^{(m,n)} =& \sum_{k=1}^{N_\text{seg}} \sum_{l=1}^{k} \Omega_k^{(m)} \Omega_l^{(n)} \int_{(k-1)\tau/N_\text{seg}}^{k\tau/N_\text{seg}} dt_2 \int_{(l-1)\tau/N_\text{seg}}^{\min(t_2,l\tau/N_\text{seg})} dt_1
\nonumber \\
& \bigg[ -\sum_{p=1}^{N} 4 \eta_{p}^{(m)} \eta_{p}^{(n)} \sin[\omega_p(t_2-t_1)]\cos(\mu t_1)\cos(\mu t_2)\bigg] 
\nonumber \\ 
=& (\vec{\Omega}^{(m)})^T \hat{D}^{(m,n)} \vec{\Omega}^{(n)}
\nonumber \\ 
=& \left\{ \begin{array}{lr} \theta^{(m,n)} & \text{if $m$ and $n$ are to be entangled,} \\ 0 & \text{otherwise,} \end{array} \right.
\end{align}
\normalsize
where $\hat{D}^{(m,n)} = \hat{D}^{(n,m)}$ is the triangular matrix with elements that are the segmented double integrals and $\theta^{(m,n)}$ denotes the desired degree of entanglement between the qubit pair ($m,n$). We note that, according to \eq{Evol}, the desired evolution to be induced between qubits $m$ and $n$ is $\exp[-i(\chi^{(m,n)}+\chi^{(n,m)}) \sigma_x^{(m)} \sigma_x^{(n)}/4]$. Since the $\chi$s are scalars, $\chi^{(m,n)}+\chi^{(n,m)} = \chi^{(m,n)}+(\chi^{(n,m)})^T$. Therefore, the constraint \eq{condB1} may be rewritten as
\small
\begin{align}
\label{eq:condB}
&(\vec{\Omega}^{(m)})^T \hat{S}^{(m,n)} \vec{\Omega}^{(n)}  \nonumber \\
&= \left\{ \begin{array}{lr} \theta^{(m,n)} & \text{if $m$ and $n$ are to be entangled,} \\ 0 & \text{otherwise,} \end{array} \right.
\end{align}
\normalsize
where $\hat{S}^{(m,n)} = [\hat{D}^{(m,n)} + (\hat{D}^{(m,n)})^T]/2$ is a symmetric matrix. The problem of finding the amplitude vectors $\vec{\Omega}$ satisfying the two conditions \eq{condA} and \eq{condB} can in principle be written in the form of a quadratically constrained quadratic program (QCQP) \cite{QCQP}, which is in general NP-hard, as has been pointed out in the literature \cite{ar:Caroline,ar:Kim}. However, our problem is fully specified by the two equations \eq{condA} and \eq{condB}, which is a special case of QCQP. The vectors $\vec{\Omega}$ that satisfy \eq{condA} and \eq{condB} can be solved exactly in polynomial time using a linear approach.

\section{EASE gate protocol}

\begin{figure*}[ht!]
\centering
\includegraphics[scale=0.4]{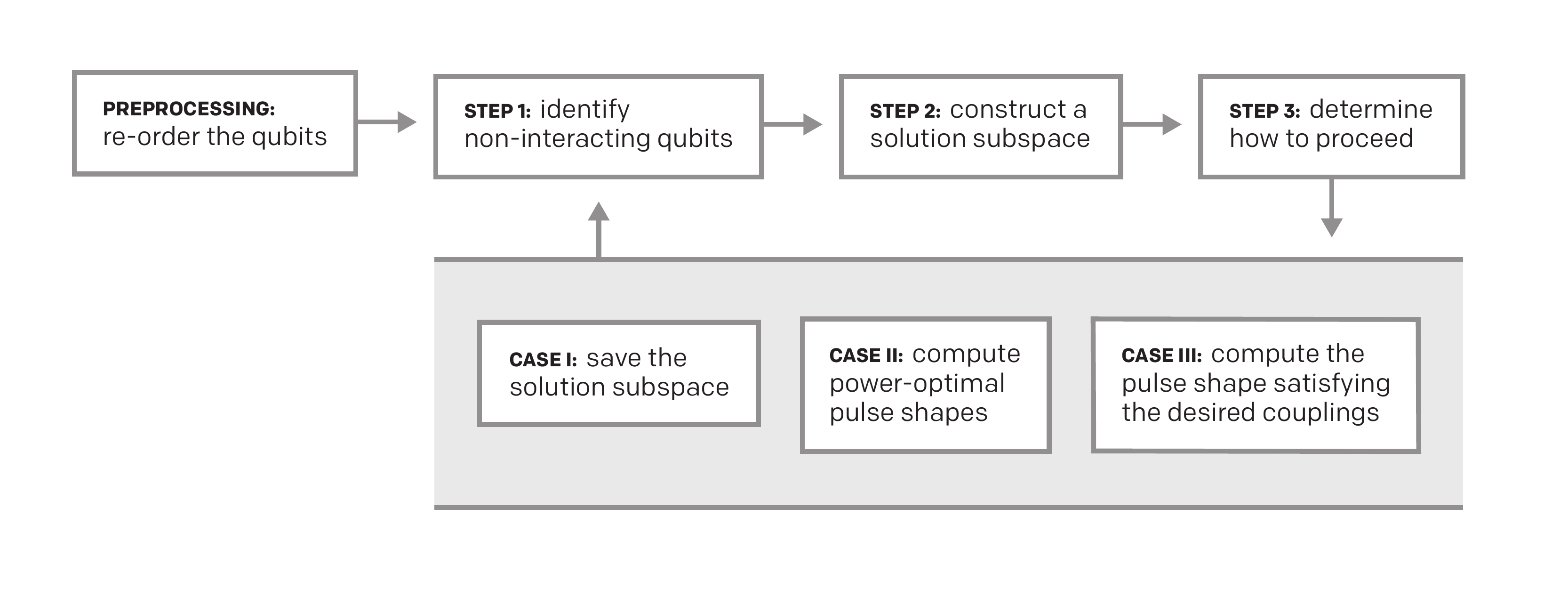} 
\caption{\textbf{Flowchart for EASE-gate pulse shape synthesis.} As a preprocessing step, reorder those qubits that participate in a given EASE gate into disjoint, non-interacting sets, where the first and second qubits of each set interact. Consider now the following iterative steps. In Step 1, identify the qubits considered in the past iterations that do not interact with the currently considered qubit. In Step 2, construct a subspace orthogonal to the interactions between previously determined pulse shapes for qubits identified in Step 1 and the currently considered qubit. In Step 3, determine how to proceed based on the index of the currently considered qubit within its set. If that qubit is the first element of its set, proceed to Case I; if it is the second, proceed to Case II; otherwise, proceed to Case III. Case I: save the orthonormalized vectors spanning the subspace constructed in Step 2. Some linear combination of those vectors will yield a power-optimal pulse shape for the currently considered qubit. Case II: using two sets of orthonormal vectors, for the currently considered and the immediately preceding qubits, compute the power-optimal pulse shapes for those qubits given their interaction matrix. Case III: 
Compute the pulse shape for the currently considered qubit that satisfies the desired entangling interactions between itself and all of the previously considered qubits. 
We iterate Steps 1--3 until all participating qubits have been accounted for. See SOM section S2 for further details.}
\label{fig:flowchart}
\end{figure*}
Figure 2 shows a flowchart that outlines our linear approach to produce pulse shapes that implement an EASE gate. Once the experimental parameters, such as the number and positions of the ion qubits, the motional mode frequencies of the ion chain, the Lamb-Dicke parameters, the detuning frequency, the desired EASE gate duration, the number of AM segments, and the qubit pairs with corresponding degrees of entanglement are specified, our protocol constructs the $\hat{M}$-matrix in \eq{condA}. The null-space vectors of $\hat{M}$ are then computed. They span a vector space from which we draw pulse shapes that satisfy \eq{condB}. 

To find a suitable pulse shape that requires minimal laser power, an important experimental concern, the $\hat{S}$ matrix in \eq{condB} is first projected onto the null space of $\hat{M}$. The eigenvector $\vec{c}$ with the largest absolute eigenvalue of the projected matrix is then guaranteed to require the minimal power possible, measured according to the sum of squares of the individual amplitudes $\Omega^{(m)}_l$. This methodology can then be iterated to find the pulse shapes for all ion qubits involved in the EASE gate [see supporting online material (SOM) sections S1 and S2 for theoretical details] by considering the pulse-shape search-space for a given qubit to be the intersection between the full null space and a subspace orthogonal to the space of previously identified pulse shapes for ions that the given qubit needs to be decoupled from.

For an EASE operation with $N_\text{EASE}$ participating qubits, we require only $N_\text{seg} = 2N+N_\text{EASE}-1$ as the minimal number of segments. We note that, even though an EASE gate may require as many as $N_\text{EASE}(N_\text{EASE}-1)/2$ angle parameters, $N_\text{EASE}-1$ additional segments (to $2N$ segments, necessary to fulfill the null-space requirement) are sufficient to satisfy all $\chi^{(m,n)}$ relations. This is enabled by the fact that, for every additional participating qubit, we may start with the full null-space vectors that always satisfy condition (A), and thus the sets of relations associated with each qubit, according to condition (B), are independent from one another. Because our approach is completely linear, the EASE-gate pulse-shapes that exactly implement the desired operation are obtained in polynomial time.

\section{Implementation}

We implement our EASE-gate protocol on a TIQIP hosted at IonQ \cite{system2}, which can load and control small chains of ${}^{171}{\rm Yb}^+$ ion qubits. Each qubit is optically initialized to a pure quantum state and then manipulated by addressing the qubit with pulses from a mode-locked 355 nm pulsed laser. These pulses can be engineered to drive either single-qubit operations by coupling to the internal (spin) degree of freedom of the ion, or two-qubit operations by coupling to both the internal and external (collective motional) degrees of freedom. We realize EASE gates by coupling the internal and external degrees of freedom of many ions simultaneously with segmented AM laser pulses.

\begin{figure*}[t!]
\centering
\includegraphics[keepaspectratio,width=18cm]{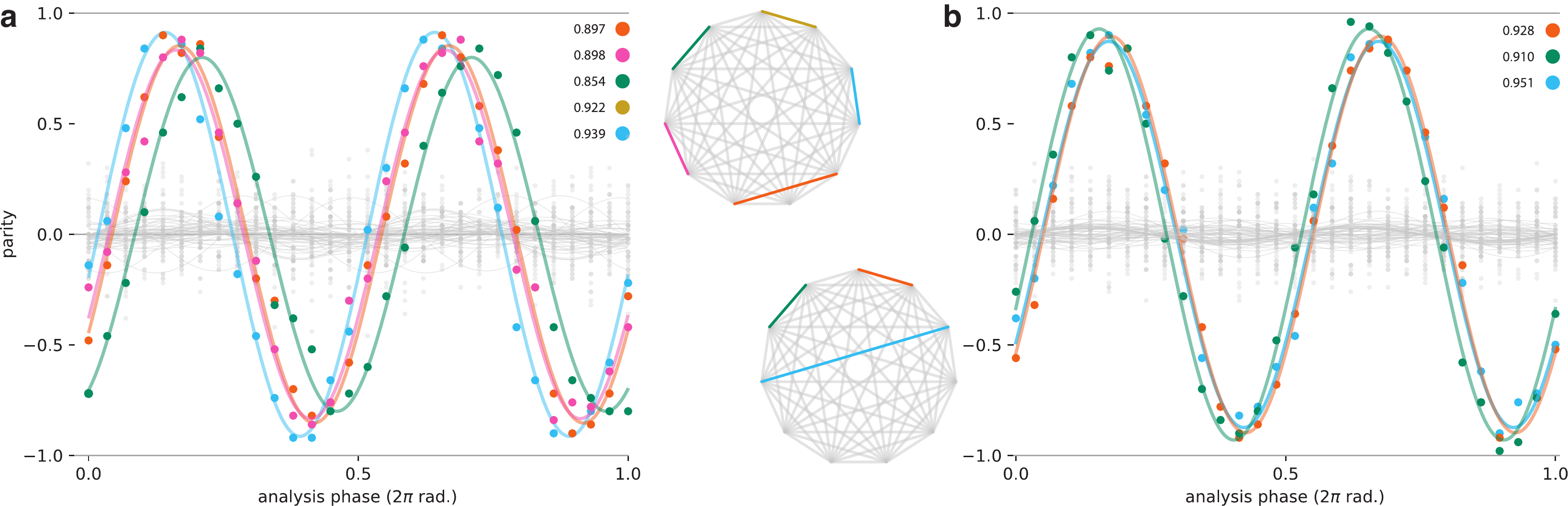} 
\caption{\textbf{Parity curves for EASE gates, and 11-qubit TIQIP all-to-all connectivity diagrams displaying the ion pairs used in the experiments.} The associated fidelities are computed from the amplitudes of the measured parity and populations via maximum likelihood estimation. \textbf{a} Parity curve for an EASE gate with 5 simultaneous XX interactions. We chose pulses with $N_\text{seg}$ = 43 and gate time $\tau$ = 816 $\mu$s. This should be compared to a single two-qubit gate time of $\tau_{min} \geq$ 300 $\mu$s; therefore, to create the same final state of all qubits, we save approximately 684 $\mu$s, 46$\%$ of the total gate time needed. This gate yielded an average fidelity of $90.2^{+0.8}_{-1.2}\%$ with an average crosstalk error of $4.8^{1.9}_{-1.2}\%$. \textbf{b} Parity curve for an EASE gate with 3 simultaneous XX interactions. We chose pulses with $N_\text{seg}$ = 35 and gate time $\tau$ = 787 $\mu$s, which  yielded an average fidelity of $93.0^{+0.6}_{-1.1}\%$ with an average crosstalk error of $6.4^{+1.5}_{-1.4}\%$. The quoted errors are 1$\sigma$ confidence intervals from the maximum likelihood estimation.}
\label{fig:parity}
\end{figure*}

In particular, we implemented EASE gates to fully entangle qubits in multiple disjoint pairs in a system with 11 ion-qubits. Of these qubits, up to 5 pairs (10 qubits) were simultaneously entangled. We then performed partial output state tomography on each entangled state by measuring the parity of the entangled pairs as a function of an analysis-pulse angle (shown in \fig{parity}), and also measuring the even parity population without applying analysis pulses. By extracting the amplitude of the measured parity and populations via maximum likelihood estimation \cite{Ballance2016,system2}, we are able to get a lower-bound estimate of the fidelity of the performed EASE gate. For our implementation with five simultaneous gates (\fig{parity}a), we estimate an average gate fidelity of $F=90.2^{+0.8}_{ -1.2}\%$. For the case in which we applied three simultaneous gates (\fig{parity}b), we estimate an average gate fidelity of $F=93.0^{+0.6}_{ -1.1}\%$. The given errors on fidelity represent a 1$\sigma$ confidence interval on the maximum likelihood estimation used to determine the fidelity.

We use the same technique to estimate any residual entanglement with non-addressed ions, due predominantly to optical crosstalk, by determining the overlap of any pair with the fully entangled bell-state we are trying to prepare. For pairs with one ion participating in a gate, the fidelity is ideally $F=25\%$, which corresponds to a fully mixed state. For pairs where neither ion participates in an applied gate, we expect to have $F=50\%$ because the initial pure state has 50$\%$ overlap with the bell state we are trying to prepare. The 50 non-involved pairs have $\delta F=4.8^{+1.9}_{-1.2}\%$ average deviation from the ideal fidelity for the five simultaneously applied gates (\fig{parity}a). In the case of three simultaneously applied gates (\fig{parity}b), we see an average deviation from the ideal fidelity of $\delta F=6.4^{+1.5}_{-1.4}\%$. In these results, we have performed more simultaneous XX gates than previously reported \cite{ar:Caroline} on chains of ions at least twice as long as any previously reported results \cite{ar:Kim, ar:Caroline}. The fidelities reported here are markedly lower; however, it should be noted that our results are not corrected for state-preparation and measurement errors. 


\section{Discussion}

\begin{figure*}[t!]
\centering
\includegraphics[scale=1.4]{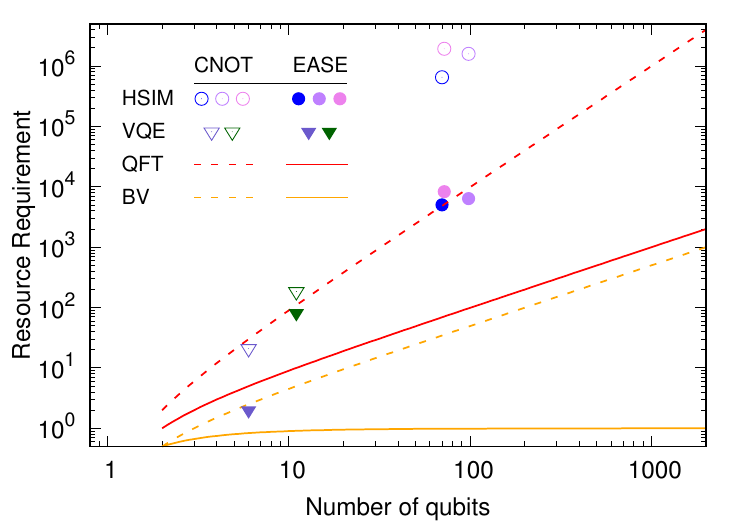} 
\caption{\textbf{Quantum resource requirement as a function of the number of qubits for various algorithms.} The resources are counted as the number of two-qubit CNOT gates for non-EASE-based implementation and multi-qubit EASE gates for EASE-based implementation. Shown are the Hamiltonian simulation (HSIM) algorithms simulating the Heisenberg Hamiltonian over various connectivity structures \cite{ar:HINT}, variational quantum eigensolver (VQE) circuits simulating the water molecule with varying degrees of approximations \cite{ar:VQE}, quantum Fourier transform (QFT) circuits \cite{ar:GlobalMS}, and Bernstein-Vazirani (BV) algorithm \cite{ar:BV} with expected gate counts over all possible oracles of a fixed size. Quadratic improvements in the resource requirement are observed for HSIM and QFT, and a linear to constant complexity improvement is observed for the BV and the Hidden-shift (not shown) algorithms.
See SOM section S3 for details on how to obtain EASE-gate counts.
}
\label{fig:improvement}
\end{figure*}


Because a TIQIP can induce couplings between arbitrary pairs of qubits by simply switching on and off pairwise interactions, the EASE gates developed and demonstrated here can readily be implemented on a TIQIP through software alone. This is in contrast to other quantum hardware platforms such as a solid-state QIPs, where each two-qubit interaction has to be hard-wired during the manufacturing process. TIQIPs can load as many qubits as necessary and employ the EASE-gate protocol to simultaneously implement any combinations of simultaneously addressible Ising interactions with little to no extra cost at the hardware level.


A host of quantum algorithms benefit from the ability to implement EASE gates. 
These algorithms tend to contain an orderly structure such that the circuit may be manipulated to reveal multiple Ising interactions applied simultaneously.
For instance:
\begin{itemize}
\item Quantum arithmetic circuits \cite{ar:Beauregard,ar:Kutin} -- useful for solving an integer factoring problem or computing discrete logarithms over Abelian groups \cite{ar:Shor}.
\item Multi-control Toffoli gates using global XX gates as a special instance of an EASE gate \cite{ar:GlobalMS} -- useful for e.g., Grover's unsorted database search algorithm \cite{ar:Grover}, applicable for solving certain satisfiability problems \cite{ar:IEEE}.
\item Fan-in or fan-out CNOTs or various roots of NOTs -- useful for realizing the quantum Fourier transform \cite{ar:GlobalMS} or the Bernstein-Vazirani algorithm \cite{ar:BV}.
\item Disjoint $k$-local operators -- useful for quantum simulation circuits, including both variational quantum eigensolver \cite{ar:VQE} or Hamiltonian-dynamics simulations \cite{ar:HINT}, and the Hidden-shift algorithms \cite{ar:HS}.
\end{itemize}
To highlight the advantages offered by the EASE operation, in \fig{improvement} we show a selection of notable algorithms that benefit from our efficient EASE-gate protocol.


Our EASE-gate protocol is linear and the pulse shapes we obtain exactly solve the problem and induce the desired quantum operation with up to $N(N-1)/2$ angle parameters $\theta^{(m,n)}$ with minimal control overhead, i.e., linear in $N$, comparable to a single XX gate in terms of the number of segments. The shapes are generated in time polynomial in the system size and are power-optimal for the AM approach when used for a single XX gate. This is in contrast to the non-linear, approximate methods used in previous studies \cite{ar:Caroline,ar:Kim} that in general return an approximate pulse-shape solution and require an exponential overhead in the number of segments. Our protocol explains why it was possible in previous studies \cite{ar:Caroline} that a certain echo-based pulse-shape ansatz worked well for applying simultaneous gates on disjoint pairs of qubits -- the shape automatically satisfies the entanglement requirement condition (B) and the infidelity owing to the imperfect decoupling from the motional modes, due to condition (A), may be minimized by navigating through the null space of $\hat{M}$. Furthermore, our protocol enables us to entangle pairs of qubits with overlapping qubits.
Our protocol is scalable and is guaranteed to work for any modulation that admits a linear construction, such as ours. 

We note that other quantum information processor architectures, such as those based on quantum dots \cite{qdot}, neutral atoms, \cite{rydberg1,rydberg2} or superconducting circuits \cite{super1,super2}, also employ pulse-shape techniques to induce desired quantum operations. While the evolution operators for these approaches are not identical to the one considered here, the motivation behind the pulse shaping is the same: Remove the unwanted coupling while preserving the desired interaction from the architecturally-inducible Hamiltonian. We anticipate that the kind of efficient, linear approach we show here may be applicable for other qubit technologies with further research.

\section{Outlook}
Classical supercomputers employ Multi-Instruction Multi-Data architectures and today's personal computers typically employ Single-Instruction Multi-Data architectures. These parallel architectures have contributed significantly to sustaining the growth of classical processing power in the era where the frequency scaling of the processors has halted. Likewise, we expect the EASE protocol we explore in this paper to significantly boost the power of quantum computing, unlocking its ability to implement multiple entangling gates efficiently. Akin to the well-known Amdahl's law in classical parallel computing \cite{ar:Amdahl}, we may roughly estimate the quantum latency to scale inversely proportional to $1-p+2p/N^2$, where $p$ denotes the proportion of the quantum computational task that benefits from the simultaneous operations and the factor $N^2/2$ arises from the capability of the EASE gate to implement up to ${\approx} N^2/2$ entangling gates at a time. We believe simultaneously entangling gates, such as the EASE gates developed in this paper, will help ensure continued growth of the power of quantum processors, even when we encounter resource limitations per qubit.

\section*{Acknowledgement}
The authors would like to thank all members of the IonQ, Inc. team, specifically the efforts of Coleman Collins, Jonathan Mizrahi, Kai Hudek, and Jamie David Wong-Campos for design and implementation of experimental system components, as well as assistance with interpretation and visual illustrations of figures presented here. 

\section*{Author contributions}
R.B., Y.N, and N.G. devised the linear null-space 
method for the efficient construction of pulses. N.G. devised the EASE protocol under Y.N.'s supervision. N.G., R.B. and Y.N. carried out the in-silico implementation of the protocol. The apparatus was designed and built by K.B., K.W., V.C., J.A., S.D., N.P., and JS.C. and experimental data was collected and analyzed by K.B. and K.W.; Y.N., K.W., R.B., N.G., and K.B. prepared the manuscript with input from all authors. 



\section*{Supporting Online Material}

\section*{MATERIALS AND METHODS}
\subsection*{S1. Single XX-gate}
As discussed thoroughly the literature \cite{ar:Caroline,ar:Kim,AM1,AM2,FM_umd}, the XX gate on an $N$-ion chain is implemented by ensuring the residual motion $\alpha_p^{(m)}$ for each ion $m$ and motional mode $p$ is zero, while the effective spin-spin interaction $\chi^{(m,n)}$ for the ion pair ($m$, $n$) is non-zero. Specifically, in an amplitude-modulated implementation where we segment the pulse $\Omega^{(m)}(t)$ applied to the $m$th ion into $N_\text{seg}$ equi-spaced segments, we require
\begin{equation}
\label{eq:condition_null}
\hat{M}\vec{\Omega}^{(m)} = \vec{0},
\end{equation}
where the matrix elements of $\hat{M}$ are
\begin{align}
\label{eq:matrixm}
\hat{M}_{p,k} =  \int_{\tau_{k-1}}^{\tau_{k}} \cos(\mu t) \cos(\omega_p t) dt, \nonumber \\
\hat{M}_{p+N,k} =  \int_{\tau_{k-1}}^{\tau_{k}} \cos(\mu t) \sin(\omega_p t) dt,
\end{align}
and 
\begin{equation}
\vec{\Omega}^{(m)}  = (\Omega_1^{(m)} \; \Omega_2^{(m)} \; ... \; \Omega_{N_\text{seg}}^{(m)})^T,
\end{equation}
where $\Omega_{k}^{(m)}$, $k=1,2,...,N_\text{seg}$, is the signed amplitude of the $k$th segment, $\mu$ is the detuning from the carrier frequency, $\omega_p$ is the mode frequency, and $\tau_k = k\tau/N_\text{seg}$. Additionally, we require
\begin{equation}
\label{eq:condition_chi}
\chi^{(m,n)} = (\vec{\Omega}^{(n)})^T \hat{S}^{(m,n)} \vec{\Omega}^{(m)} \neq 0
\end{equation}
with the symmetric matrix
\begin{equation}
\label{eq:smatrix}
\hat{S}^{(m,n)} = [\hat{D}^{(m,n)} + (\hat{D}^{(m,n)})^T]/2,
\end{equation}
where the matrix elements of the triangular matrix $\hat{D}^{(m,n)}$ are
\small
\begin{align}
 \hat{D}^{(m,n)}_{k,l} = 
& \int_{\tau_{k-1}}^{\tau_k} dt_2 \int_{\tau_{l-1}}^{\min(t_2,\tau_l)} dt_1
 \nonumber \\
& \bigg[ - \sum_{p=1}^{N} 4 \eta_{p}^{(m)} \eta_{p}^{(n)} \sin[\omega_p(t_2-t_1)]\cos(\mu t_1)\cos(\mu t_2)\bigg] 
 \end{align}
 \normalsize
and $\eta_p^{(m)}$ is the Lamb-Dicke parameter. 

To satisfy conditions \eq{condition_null} and \eq{condition_chi}, we require $N_\text{seg} > 2N$, such that there is at least one dimension available in the null-space of $\hat{M}$. We then linearly combine the orthonormal null-space vectors $\vec{\Omega}_\text{null}^{[i]}$, where $i=1,2,...,{\mathcal N}$, where ${\mathcal N}$ is the dimension of the null space of $\hat{M}$, to find a suitable vector $\vec{\Omega}^{(m)}$. We further assume that $\vec{\Omega}^{(m)} = \vec{\Omega}^{(n)}$, i.e., we illuminate the two qubits $m$ and $n$ targeted by the XX gate with the same pulses. Using now the ${\mathcal N}$ degrees of freedom, we may at this point optimize with respect to certain experimentally favorable conditions, such as the laser power. We found that, while the condition translates to finding the vector $\vec{\Omega}^{(m)}$ which minimizes $\text{max}_l \Omega_l^{(m)}$, approximating this condition to finding the smallest sum of squares of $\Omega_l^{(m)}$ works well.

Specifically, to find the approximate solution to the power minimization requirement, we first reduce the solution space characterized by $\hat{S}^{(m,n)}$ to within the null space of $\hat{M}$. We do this by conjugating $\hat{S}^{(m,n)}$ with the orthonormal null-space vectors $\vec{\Omega}_\text{null}^{[i]}$ for $i=1,2,..,{\mathcal N}$, i.e., we construct the reduced matrix $\hat{V}^{(m,n)}$ with matrix elements
\begin{equation}
\label{eq:vmatrix}
\hat{V}^{(m,n)}_{i,j} = (\vec{\Omega}_\text{null}^{[i]})^T \hat{S}^{(m,n)} \vec{\Omega}_\text{null}^{[j]}.
\end{equation}
We then find the normalized eigenvector $\vec{c}$ of $\hat{V}^{(m,n)}$ with the largest absolute eigenvalue $\lambda$ . The eigenvector $\vec{c}$ may now be used to find the desired pulse:
\begin{equation}
\vec{\Omega}^{(m)} = \vec{\Omega}^{(n)} = \frac{\theta^{(m,n)}}{\lambda} \sum_{i=1}^{\mathcal N} c_i \vec{\Omega}_\text{null}^{[i]}.
\end{equation}

\subsection*{S2. EASE gate}
To find pulses that entangle multiple pairs of ions simultaneously, we once again start with the full set of null-space vectors $\vec{\Omega}_{\text{null}}^{[i]}$, $i=1,2,..,{\mathcal N}$. These vectors, by construction, automatically decouple the spin and motional states and satisfy the condition stated in \eq{condition_null}. Therefore, for any qubits that participate in a given EASE gate, the search for a suitable pulse starts from the null space spanned by $\vec{\Omega}_{\text{null}}^{[i]}$. 

Consider now an EASE gate that operates on $N_\text{EASE}$ qubits. As a preprocessing step, we first perform the following:

{\it Preprocessing} -- Reorder the qubit indices such that disjoint sets of qubits are grouped and labeled consecutively; in a graph constructed with vertices that denote each qubits and edges that denote constituent XX gates of a given EASE gate with non-zero degree of entanglement, a pair of qubits may be considered disconnected if there is no path in this graph that has the two qubits as endpoints. A collection of connected qubits form a disjoint set. In particular, the ordering is done for each of the disjoint sets in such a way that the first and second element of each set are directly connected, i.e., there is an edge that connects the two qubits.

We now find a suitable pulse for each of the $N_\text{EASE}$ participating qubits. The following procedure may then be iterated for each ion to obtain the desired pulse.

{\it Step 1} -- For the $n$th qubit, identify the desired XX interaction strengths between the previous $n{-}1$ qubits and the current $n$th qubit. Then, collect those interactions whose strengths are zero, i.e., determine which of the $n{-}1$ qubits are not to be coupled with the $n$th qubit.

{\it Step 2} -- From the full null space spanned by $\vec{\Omega}_{\text{null}}^{[i]}$, construct a subspace that is orthogonal to the vectors $\hat{S}^{(m,n)}\vec{\Omega}^{(m)}$, where $m{<}n$ are the qubit indices of those qubits not coupled to the $n$th qubit, identified in Step 1.

{\it Step 3} -- Determine which of the following three cases are applicable. 
Case I -- No qubits $m{<}n$ are coupled to the $n$th qubit. This case occurs for every first qubit in a distinct disjoint set.
Case II -- Some qubits $m{<}n$ are to be coupled to the $n$th qubit, but one of them, e.g., qubit index $n{-}1$, has a yet-to-be-determined pulse shape. This case occurs for every second qubit in a distinct disjoint set.
Case III - Some qubits $m{<}n$ are to be coupled to the $n$th qubit and the pulse shapes for all of those qubits are already determined. This case occurs for the third qubit and onward in a distinct disjoint set.

The three cases I--III determine which procedure we select as the final step of the iteration before the start of another iteration. In particular, 
\begin{itemize}
\item Case I -- Save the orthonormal vectors $\vec{v}^{(n)}_l$, $l=1,2,...,{\mathcal N}-(n-1)$, that span the subspace identified in Step 2. Continue to qubit number $n{+}1$.
\item Case II -- Compute orthonormal vectors $\vec{v}^{(n)}_l$, $l=1,2,...,{\mathcal N}-(n-2)$, that span the subspace identified in Step 2. Also compute $\hat{V}_{i,j}^{(n-1,n)} = (\vec{\Omega}_\text{null}^{[i]})^T \hat{S}^{(n{-}1,n)} \vec{\Omega}_\text{null}^{[j]}$ and find the eigenvector $\vec{c}$ with the largest absolute eigenvalue. The desired pulse shape solutions for qubit $n{-}1$ and $n$ are the closest pulse shapes that can be generated from their respective pulse-shape search-space to $\sum_i c_i \vec{\Omega}_\text{null}^{[i]}$. Note that the vectors $\vec{v}^{(n{-}1)}$ span the pulse-shape search-space for the $n{-}1$st qubit. Once the pulse shapes for both qubits, $n{-}1$ and $n$, are determined, continue to qubit number $n{+}1$. 
\item Case III -- Restarting from Step 1, i.e., with all of the null-space vectors $\vec{\Omega}_\text{null}^{[i]}$, compute the basis vectors that span the space, in which the inner-product relations with $\hat{S}^{(m,n)}\vec{\Omega}^{(m)}$ for all qubits $m{<}n$ are satisfied. The inner products here denote the coupling strengths. The pulse shape for the $n$th ion is an appropriate combination of the basis vectors such that the norm is minimized. Continue to qubit number $n{+}1$.
\end{itemize}

We next detail the computational steps used to address each case. For case I, we need to find orthonormal vectors $\vec{v}^{(n)}$ drawn from the full null space spanned by $\vec{\Omega}_{\text{null}}^{[i]}$ such that they are orthogonal to $\hat{S}^{(m,n)}\vec{\Omega}^{(m)}$, $m=1,2,..,n{-}1$. Therefore, we require that
\begin{align}
(\vec{\Omega}^{(m)})^T \hat{S}^{(m,n)} \vec{\Omega}^{(n)}
= (\hat{V}^{(m,n)} \vec{\xi}^{(m)})^T \vec{\zeta}^{(n)} = 0,
\end{align}
where we used $\hat{V}^{(m,n)}$ in \eq{vmatrix}, $\vec{\Omega}^{(m)} = \sum_i \xi^{(m)}_i \vec{\Omega}_\text{null}^{[i]}$, and $\vec{\Omega}^{(n)} = \sum_i \zeta_i^{(n)} \vec{\Omega}_\text{null}^{[i]}$. To find $\vec{\zeta}^{(n)}$, we define a matrix $\hat{\epsilon}$ whose rows are $(\hat{V}^{(m,n)} \vec{\xi}^{(m)})^T$. We then obtain $\hat{\tilde{\epsilon}} \vec{\zeta}^{(n)} = \vec{0}$, where $\hat{\tilde{\epsilon}}$ is the row-reduced $\hat{\epsilon}$. This means that
\begin{equation}
\zeta_{j'}^{(n)} = -\sum_{j=n}^{\mathcal N} \tilde{\epsilon}_{j',j} \zeta_j^{(n)}
\end{equation}
for $j'=1,2,..,n-1$,
and thus the desired $\zeta^{(n)}$-space is spanned by vectors $\vec{\rho}_l$, where
\begin{align}
\label{eq:rhovector}
\tiny
\vec{\rho}_l \in \left\{ 
\left(\begin{matrix} -\tilde\epsilon_{1,n} \\  -\tilde\epsilon_{2,n} \\ \vdots \\ -\tilde\epsilon_{n-1,n} \\ 1 \\ 0 \\ 0 \\ \vdots \\ 0 \end{matrix}\right), 
\left(\begin{matrix} -\tilde\epsilon_{1,n+1} \\  -\tilde\epsilon_{2,n+1} \\ \vdots \\ -\tilde\epsilon_{n-1,n+1} \\ 0 \\ 1 \\ 0 \\ \vdots \\ 0 \end{matrix}\right),
\cdots,  
\left(\begin{matrix} -\tilde\epsilon_{1,{\mathcal N}} \\  -\tilde\epsilon_{2,{\mathcal N}} \\ \vdots \\ -\tilde\epsilon_{n-1,{\mathcal N}} \\ 0 \\ 0 \\ 0 \\ \vdots \\ 1 \end{matrix}\right)
\right\}.
\end{align}
To this end, we obtain $\vec{v}^{(n)}_l$ in Case I to be $\sum_i (\vec{\varrho}_l)_i \vec{\Omega}_\text{null}^{[i]}$, where $\vec{\varrho}_l$ are the orthonormalized $\vec{\rho}_l$ in \eq{rhovector}.

For case II, we start with
\begin{align}
&(\vec{\Omega}^{(n{-}1)})^T \hat{S}^{(n{-}1,n)} \vec{\Omega}^{(n)} \nonumber \\
&= \left[\sum_{l=1}^{n-2}b_l \left(\sum_{i=1}^{\mathcal N} \left(\vec{\varrho}_l^{(n-1)}\right)_i \vec{\Omega}_\text{null}^{[i]}\right)\right]^T  \nonumber \\ &\qquad \hat{S}^{(n{-}1,n)} \left[\sum_{l'=1}^{n-2}b'_{l'} \left(\sum_{i'=1}^{\mathcal N} \left(\vec{\varrho}_{l'}^{(n)}\right)_{i'} \vec{\Omega}_\text{null}^{[i']} \right)\right] 
\nonumber \\
&= \left[\sum_{i=1}^{\mathcal N} \left(\sum_{l=1}^{n-2} b_l \left(\vec{\varrho}_l^{(n-1)}\right)_i\right)\vec{\Omega}_\text{null}^{[i]}\right]^T \nonumber \\ &\qquad \hat{S}^{(n{-}1,n)}
\left[\sum_{i'=1}^{\mathcal N} \left(\sum_{l'=1}^{n-2} b'_{l'} \left(\vec{\varrho}_{l'}^{(n)}\right)_{i'}\right)\vec{\Omega}_\text{null}^{[i']}\right]
\nonumber \\
&= \vec{B}^T \hat{V}^{(n{-}1,n)} \vec{B}',
\end{align}
where $B_i = \sum_l b_l (\vec{\varrho}_l^{(n-1)})_i$ and $B'_{i'} = \sum_{l'} b'_{l'}(\vec{\varrho}_{l'}^{(n)})_{i'}$. Denoting the eigenvector of $\hat{V}$ with the largest absolute eigenvalue as $\vec{L}$, we aim to find $\vec{B}$ and $\vec{B}'$ that have the largest overlap with $\vec{L}$. We obtain the suitable choices of $b_l$ by computing the inner product between $\vec{\varrho}_l^{(n-1)}$ and $\vec{L}$, and likewise for $b'_{l'}$; in particular, $b_l = \vec{\varrho}_l^{(n-1)}\cdot \vec{L} / A$ and $b'_{l'} = \vec{\varrho}_{l'}^{(n)}\cdot \vec{L} / A'$, where the normalization constants $A$ and $A'$ are chosen such that $\vec{B}^T \hat{V}^{(n{-}1,n)} \vec{B}' = \theta^{(n{-}1,n)}$.

For case III, we aim to find $\vec{\Omega}^{(n)}$ such that the main text Eq.~(4) 
is satisfied. To do so, we start with all of the null-space vectors $\Omega_\text{null}^{[i]}$. Then, we require that
\begin{align}
\label{eq:case3conditions}
(\vec{\Omega}^{(m)})^T \hat{S}^{(m,n)} \vec{\Omega}^{(n)}
= (\hat{V}^{(m,n)} \vec{\xi}^{(m)})^T \vec{\zeta}^{(n)} = \chi^{(m,n)},
\end{align}
where $\chi^{(m,n)}$ denotes the degree of entanglement between qubits $m$ and $n$, $m {<} n$, $\hat{V}^{(m,n)}$ is defined in \eq{vmatrix}, $\vec{\Omega}^{(m)} = \sum_i \xi^{(m)}_i \vec{\Omega}_\text{null}^{[i]}$, and $\vec{\Omega}^{(n)} = \sum_i \zeta_i^{(n)} \vec{\Omega}_\text{null}^{[i]}$.
To find $\vec{\zeta}^{(n)}$, we once again define a matrix $\hat{\epsilon}$ whose rows are $(\hat{V}^{(m,n)} \vec{\xi}^{(m)})^T$. We then obtain $\hat{\tilde{\epsilon}} \vec{\zeta}^{(n)} = \vec{\tilde{\chi}}^{(n)}$, where $\hat{\tilde{\epsilon}}$ is the row-reduced $\hat{\epsilon}$ and the vector elements $\tilde{\chi}_m^{(n)}$ are the accordingly row-operated $\chi^{(m,n)}$ values. Then, the subspace that satisfies the aforementioned inner-product conditions is spanned by
\begin{equation}
\left(\vec{\rho}_l\right)_i = 
\begin{cases}
\tilde{\chi}_i^{(n)} - h_{n+l-1}\tilde{\epsilon}_{i,n+l-1} \quad &\text{if } i < n,\\
h_{n+l-1} \quad &\text{if } i = n+l-1,\\
0 \quad &\text{otherwise,}
\end{cases}
\end{equation}
where $h_{n+l-1}$ are free parameters and $l=1,2,...,{\mathcal N}-(n-1)$.
\comment{
\begin{equation}
\label{eq:rhovector}
\tiny
\vec{\rho}_l \in \left\{ 
\left(\begin{matrix} \tilde{\chi}_1^{(n)}-c_{Q+k+1}\tilde\epsilon^{[q]}_{1,Q+k+1} \\  \tilde{\chi}_2^{(n)}-c_{Q+k+1}\tilde\epsilon^{[q]}_{2,Q+k+1} \\ \vdots \\ \tilde{\chi}_Q^{(n)}-c_{Q+k+1}\tilde\epsilon^{[q]}_{Q,Q+k+1} \\ c_{Q+k+1} \\ 0 \\ 0 \\ \vdots \\ 0 \end{matrix}\right), 
\left(\begin{matrix} \tilde{\chi}_1^{(n)}-c_{Q+k+2}\tilde\epsilon^{[q]}_{1,Q+k+2} \\  \tilde{\chi}_2^{(n)}-c_{Q+k+2}\tilde\epsilon^{[q]}_{2,Q+k+2} \\ \vdots \\ \tilde{\chi}_Q^{(n)}-c_{Q+k+2}\tilde\epsilon^{[q]}_{Q,Q+k+2} \\ 0 \\ c_{Q+k+2} \\ 0 \\ \vdots \\ 0 \end{matrix}\right),
\cdots,  
\left(\begin{matrix} \tilde{\chi}_1^{(n)}-c_{\text{dim}(\text{Null}(\hat{M}))}\tilde\epsilon^{[q]}_{1,\text{dim}(\text{Null}(\hat{M}))} \\  \tilde{\chi}_2^{(n)}-c_{\text{dim}(\text{Null}(\hat{M}))}\tilde\epsilon^{[q]}_{2,\text{dim}(\text{Null}(\hat{M}))} \\ \vdots \\ \tilde{\chi}_Q^{(n)}-c_{\text{dim}(\text{Null}(\hat{M}))}\tilde\epsilon^{[q]}_{Q,\text{dim}(\text{Null}(\hat{M}))} \\ 0 \\ 0 \\ 0 \\ \vdots \\ c_{\text{dim}(\text{Null}(\hat{M}))} \end{matrix}\right)
\right\}.
\end{equation}
}

We now ought to find appropriate coefficients $r_l$ such that $\vec{\zeta}^{(n)} = \sum_l r_l\vec{\rho}_l$, where $\sum_l r_l = 1$, and the norm $|\vec{\zeta}^{(n)}|$ is minimized. This amounts to solving $\partial |\vec{\zeta}^{(n)}|/\partial d_l = 0$, where $d_l = h_{n+l-1} r_l$. It can be shown straightforwardly that in matrix form this may be expressed as
\begin{equation}
\vec{d} = \hat{P}^{-1} \vec{\phi},
\end{equation}
where the matrix elements of $\hat{P}$ are 
\begin{equation}
\hat{P}_{l,l'} = \begin{cases} 
1+\sum_{q=1}^{n-1} (\tilde\epsilon_{q,n+l-1})^2 \quad &\text{if }l=l', \\
\sum_{q=1}^{n-1} \tilde\epsilon_{q,n+l-1}\tilde\epsilon_{q,n+l'-1} \quad &\text{if }l\neq l',
\end{cases}
\end{equation}
where $l,l'=1,2,..,{\mathcal N}-(n-1)$,
and the vector elements of $\vec{\phi}$ are
\begin{equation}
\vec{\phi}_l = \sum_{q=1}^{n-1} \tilde\epsilon_{q,n+l-1} \tilde\chi_q^{(n)}.
\end{equation}
The desired $\vec{\zeta}$ is thus
\begin{equation}
\zeta_i = \begin{cases} 
\tilde\chi_i^{(n)} - \sum_{l=1}^{{\mathcal N}-(n-1)} \tilde\epsilon_{i,n+l-1} d_{l}, \quad &\text{if }i< n, \\
d_{i-n+1}, \quad &\text{if }i \geq n.
\end{cases}
\end{equation}

\subsection*{S3. EASE gate counts}
We detail in this section the methods used to compute the EASE gate counts shown in \fig{improvement} of the main text. The readers are strongly encouraged to read the corresponding references cited herein for each of the following considered cases. The cases considered here are (i) Heisenberg Hamiltonian simulation circuits over various connectivity patterns \cite{ar:HINT}, (ii) water molecule simulation,  using a variational eigensolver with varying degrees of approximation \cite{ar:VQE}, (iii) quantum Fourier transform \cite{ar:GlobalMS}, (iv) Bernstein-Vazirani algorithms \cite{ar:BV}, averaged over all possible oracles of a given size, and (v) Hidden-Shift algorithms \cite{ar:HS} with inner-product function.

We used the Heisenberg-Hamiltonian simulation circuits in \cite{ar:HINT} with fourth-order product formulas as benchmarks. Specifically, the considered connectivity graphs are $(3{-}5{-}70)$, $(4{-}4{-}98)$, and $(5{-}3{-}72)$, where $(k{-}d{-}n)$ is a graph with degree $k$, diameter $d$, and number of vertices $n$. The CNOT gate counts are reported in Table~I of \cite{ar:HINT}. The EASE gate counts are computed by the following procedure.

We note that the Heisenberg Hamiltonian over two qubits is of the form $\sigma_x \sigma_x + \sigma_y \sigma_y + \sigma_z \sigma_z$. Therefore, we may order the Hamiltonian terms such that all of the $\sigma_x \sigma_x$ terms appear consecutively, and likewise for $\sigma_y \sigma_y$ and $\sigma_z \sigma_z$. In this case, each stage of the fourth-order product-formula-based approximation of the evolution operator with the reordered Hamiltonian can be shown to have 30 sets of $\sigma_x \sigma_x$, $\sigma_y \sigma_y$, or $\sigma_z \sigma_z$ interactions, 9 of which can be merged since a $\sigma_x \sigma_x$ interaction followed by another $\sigma_x \sigma_x$ interaction is nothing but the combined $\sigma_x \sigma_x$ interaction. Therefore, there are a total of 21 $\sigma_i \sigma_i$, $i \in \{x,y,z\}$ interaction sets per product-formula stage. We recall that the EASE gate can implement each interaction set simultaneously. Thus, the total number of EASE gates evaluates to the number of product-formula stages times 21. The number of stages required may be found in Equation~(4) of \cite{ar:HINT}.

For the water molecule simulations, we considered HF+7 and HF+21 cases; HF stands for Hartree-Fock as defined in \cite{ar:VQE}. The HF+7 case is the pairwise excitation, bosonic-only case with 7 such terms, where the effective evolution operator is expanded using the first-order product formula. This case can be shown to require 14 XX gates (two XX gates with continuous parameters per excitation), or 21 CNOT gates, since any unitary operation over two qubits does not require more than three CNOT gates \cite{ar:Markov}. For the EASE-based approach, we reorder the excitation operators such that each one of the two XX gates per excitation can be grouped together to then be amenable to implementation by a single EASE gate. This results in two EASE gates in total for the HF+7 case. For the remaining case of HF+21, with the three-CNOT implementation above, the CNOT count is 185. For the EASE-gate consideration, we allow for parallel implementation of excitations over disjoint sets of qubits. Careful arrangement of the excitation terms results in 81 EASE gates in total.

The $n$-qubit quantum Fourier transform requires $n(n-1)$ CNOT gates, where we used two CNOT gates per controlled-$z^a$ gate, where $z^a$ is defined according to 
\begin{equation}
z^a := \left[\begin{matrix} 1\; &0 \\ 0\; &e^{i\pi a} \end{matrix}\right].
\end{equation}
With EASE gates, each of the $n{-}1$ layers of controlled-$z^a$ rotations, $a\in\{1/2,1/4,...,1/2^{l}\}$, where $l$ is the layer number, can be implemented simultaneously. Therefore, the EASE-gate counts for the quantum Fourier transform is $n{-}1$.

The Bernstein-Vazirani algorithm \cite{ar:BV}, implemented on an $n$-qubit quantum computer has an oracle size of $n-1$. Therefore, on average, $(n{-}1)/2$ CNOT gates are required to implement the oracle. The EASE gate allows for the implementation of any non-zero bit-string oracle in a single operation.

The Hidden Shift algorithm with inner-product function \cite{ar:HS}, implemented on a $n$-qubit ($n$ even) quantum computer, requires $n$ CNOT gates. Since this circuit requires two layers of parallel CNOT gates, each with $n/2$ CNOT gates, the EASE protocol allows us to implement the $n$-qubit Hidden-Shift algorithm with only two operations regardless of the number of qubits. 

\end{document}